\documentclass[letterpaper]{JHEP3}

\newcommand{\roughly}[1]{\mathrel{\raise.3ex\hbox{$#1$\kern-0.85em
\lower1ex\hbox{$\sim$}}}}

\newcommand{\lsim}{\roughly<}
\newcommand{\gsim}{\roughly>}

\def\exd{{\hbox{d}}}

\def\ba{\begin{eqnarray}}
\def\ea{\end{eqnarray}}
\def\be{\begin{equation}}
\def\ee{\end{equation}}

\def\ssB{{\scriptscriptstyle B}}
\def\ssD{{\scriptscriptstyle D}}
\def\ssF{{\scriptscriptstyle F}}
\def\ssI{{\scriptscriptstyle I}}
\def\ssJ{{\scriptscriptstyle J}}
\def\ssK{{\scriptscriptstyle K}}
\def\ssM{{\scriptscriptstyle M}}
\def\ssN{{\scriptscriptstyle N}}
\def\ssP{{\scriptscriptstyle P}}
\def\ssR{{\scriptscriptstyle R}}
\def\ssV{{\scriptscriptstyle V}}
\def\KK{{\scriptscriptstyle KK}}

\def\G{\mathcal{G}}
\def\H{\mathcal{H}}
\def\K{\mathcal{K}}
\def\L{\mathcal{L}}
\def\O{\mathcal{O}}
\def\R{\mathcal{R}}
\def\V{\mathcal{V}}
\def\U{\mathcal{U}}

\def\nn{\nonumber}

\def\({\left(}
\def\){\right)}

\def\pref#1{(\ref{#1})}

\title{\"Uber-naturalness: unexpectedly light scalars\\
 from supersymmetric extra dimensions }

\author{
C.P. Burgess,${}^{1,2}$ Anshuman Maharana${}^3$
 and F. Quevedo${}^{3,4}$\\
$^1$ Department of Physics \& Astronomy, McMaster University,
 Hamilton ON, Canada.\\
$^2$   Perimeter Institute for Theoretical Physics,
 Waterloo ON, Canada.\\
$^3$ DAMTP/CMS, University of Cambridge, 
 Cambridge CB3 0WA, UK.\\
$^4$ Abdus Salam ICTP, Strada Costiera 11, Trieste 34014, Italy. }

\date{}

\abstract { Standard lore asserts that quantum effects generically
forbid the occurrence of light (non-pseudo-Goldstone) scalars
having masses smaller than the Kaluza Klein scale, $M_\KK$, in
extra-dimensional models, or the gravitino mass, $M_{3/2}$, in
supersymmetric situations. We argue that a hidden assumption
underlies this lore: that the scale of gravitational physics,
$M_g$, ({\it e.g} the string scale, $M_s$, in string theory) is of
order the Planck mass, $M_p = \sqrt{8\pi G} \simeq 10^{18}$ GeV.
We explore sensitivity to this assumption using the spectrum of
masses arising within the specific framework of large-volume
string compactifications, for which the ultraviolet completion at
the gravity scale is explicitly known to be a Type IIB string
theory. In such models the separation between $M_g$ and $M_p$ is
parameterized by the (large) size of the extra dimensional volume,
$\V$ (in string units), according to $M_p: M_g: M_\KK: M_{3/2}
\propto 1: \V^{-1/2}: \V^{-2/3}: \V^{-1}$. We find that the
generic size of quantum corrections to masses is of the order of
$M_\KK M_{{3/2}} / M_p \simeq M_p/\V^{5/3}$. The mass of the
lighest modulus (corresponding to the extra-dimensional volume)
which at the classical level is $M_\ssV \simeq M_p/\V^{3/2} \ll
M_{3/2} \ll M_\KK$ is thus
 stable against quantum corrections. This is possible because
the couplings of this modulus to other forms of matter in the
low-energy theory are generically {\em weaker} than gravitational
strength (something that is also usually thought not to occur
according to standard lore). We discuss some phenomenological and
cosmological implications of this observation.}

\preprint{IC/2010/013\\DAMTP-2010-31}

\begin{document}

\section{Introduction}

Light scalar fields play a disproportionate role in our search for
what lies beyond the Standard Model. On one hand, there are a
variety of reasons why scalar fields are very useful: their
expectation values provide the Lorentz-invariant order parameters
for spontaneously breaking symmetries, at least within the weakly
coupled limit that is under the best theoretical control. They are
ubiquitous in supersymmetric and extra-dimensional theories, which
are among the best motivated we have, where they arise as symmetry
partners of particles having other (4D) spins. And if they are
sufficiently light, scalars can do interesting things: they play
important roles in many of the various cosmological scenarios that
have been proposed to explain the mysteries of Cosmic Inflation
(and its alternatives), Dark Matter and/or Dark Energy.

However, scalars are notoriously difficult to keep light enough to
be relevant to present-day phenomenology. Because their masses are
difficult to protect from receiving large quantum corrections,
they are usually very sensitive to the ultra-violet (UV) sector.
For instance, a light scalar $\phi$ coupled to a heavy field
$\psi$ through a coupling $g^2 \, \phi^2 \, \psi^2$, generically
generates (see \S2\ below for more details) a loop correction to
its mass of order
\be
 \delta m_\phi^2 \simeq \left( \frac{g M}{4 \pi} \right)^2 \,,
\ee
from the graph shown in Fig.~1. Here $M$ is the mass of the heavy
$\psi$ particle, and the factors of $4\pi$ are those appropriate
to one loop (in four dimensions). Because of such contributions,
it is often only possible to obtain $m_\phi \ll g M/4\pi$ if there
is a conspiracy to cancel very precisely -- often to a great many
decimal places -- these kinds of large loop contributions against
other parameters in the underlying microscopic theory describing
the ultra-violet physics.

\centerline{
\begin{picture}(170,70)
    \thicklines
    \put(80,30){\circle{40}}
    \put(80,10){\circle*{4}}
    \multiput(77,10)(-9,0){5}{\line(1,0){4}}
    \multiput(80,10)(9,0){5}{\line(1,0){4}}
    \put(21,7){$\phi$}
    \put(135,7){$\phi$}
    \put(78,17){$\psi$}
\end{picture}}
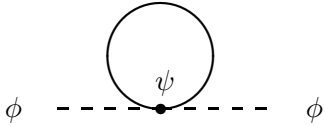
{\bf Figure 1}: A large mass correction to a light scalar from a
quartic coupling.\vspace{0.2cm}

Because of this, light scalars are rarely found in a theory's
low-energy limit, with the rare exceptions corresponding to when a
(possibly approximate) symmetry protects the scalar from receiving
large quantum corrections. On one hand, the comparative rarety of
such naturally light scalars can be regarded as a feature and not
a bug: it could explain why no fundamental scalars have yet been
found experimentally. But on the other hand, this makes it
difficult to keep scalars light enough to be useful for
understanding the origin of electroweak symmetry breaking, or to
be relevant for cosmology. It is this observation that lies at the
core of the electroweak hierarchy problem, among others.

A great deal of attention has therefore gone towards exploring
those cases where symmetries are able to protect light scalar
masses without conspiracy. The known symmetries of this kind are:
($i$) approximate shift symmetries, such as $\phi \to \phi + c$
and its nonlinear extensions, as appropriate for Goldstone and
pseudo-Goldstone bosons; ($ii$) supersymmetry, for which
cancellations between superpartners of opposite statistics
suppress contributions from heavy particles whose mass is higher
than the relevant supersymmetry-breaking scale (such as the
gravitino mass, $M_{3/2}$); and ($iii$) extra dimensions, for
which higher-dimensional symmetries (like gauge invariance or
general covariance) can protect masses from receiving quantum
corrections larger than the Kaluza-Klein (KK) scale, $M_\KK$.

Since both the supersymmetry breaking scale and the KK scale
cannot be too low without running into phenomenological
difficulties, it is usually expected that the only scalars likely
to be light enough to be relevant at very low energies are
Goldstone (or pseudo-Goldstone) bosons, despite the fact that many
candidates for fundamental theories count an abundance of scalars
among their degrees of freedom. In the absence of a protective
symmetry none of these scalars would survive to be light enough to
observe experimentally. This is true in particular for string
theory, which has both supersymmetry and extra dimensions as
sources for its many scalars.

Until recently this expectation has proven hard to test, because
of the technical difficulties associated with exploring the
spectrum of excitations near `realistic' vacua, far from the
supersymmetric configurations for which calculations are best
under control. What is new in recent times is the ability to
explore non-supersymmetric vacua to see how massive the various
would-be light scalars become once supersymmetry breaks.

Particularly interesting from this point of view are large volume
(LV) flux compactifications {\cite{LV}} of Type IIB string vacua.
These have the property that the same flux that fixes the scalar
moduli also breaks supersymmetry, producing a very predictive
spectrum of scalars whose masses depend in a predictable way on
the (large) extra-dimensional volume. Since LV models in
particular predict (at the classical level) the existence of
non-Goldstone scalars that are parametrically much lighter than
both the KK and the gravitino mass, common lore would lead one to
expect that there are large loop corrections to these classical
mass predictions, which lift their masses up to either $M_\KK$ or
$M_{3/2}$.

Our purpose in this paper is estimate the generic size of
radiative corrections to scalar masses in LV models, in order to
estimate how these compare with the predicted classical values. We
find that since loop effects are smaller than the classical
predictions, the classical masses really do provide a good
approximation to the full result. LV models therefore provide one
of the few examples of light (non-Goldstone) scalars whose masses
are naturally smaller than both the SUSY breaking and KK scales.
It turns our that their masses are nonetheless ({\em \"uber})
natural because of an interesting interplay between the scale of
supersymmetry breaking and the size of the extra dimensions.

We organize our observations as follows. First, \S2\ reviews the
generic size of one-loop corrections to scalar masses in four- and
higher-dimensional theories, in order to set conventions and
establish that our estimates reproduce standard results in
standard situations. Next, \S3\ reviews some of the properties
about the masses and couplings of scalars that arise in the LV
string vacua. This is followed, in \S4, by a discussion of how
large the radiative corrections are to these masses in LV models,
with the estimates compared with (and shown to agree with) the
results of explicit string loop calculations, when these can be
done. Finally we summarize our conclusions in \S5.

\section{Generic loop estimates}
\label{sec:GenLoopEst}

We start with a discussion of the generic size of loop effects in
four- and extra-dimensional models, contrasting how the
supersymmetric case differs from the non-supersymmetric one. This
section is meant to provide tools for later application, and
because it does not contain new material (although perhaps
presented with a slightly different point of view) it can be
skipped by the reader in a hurry.

\subsection{Technical naturalness in 4 dimensions}

To set the benchmark for what should be regarded to be a generic
quantum mass correction, consider a model involving two real
scalar fields, one of which ($\phi$) is light while the other
($\psi$) is heavy:
\be \label{phipsiL}
 -\L(\phi, \psi) :=  \frac12 \Bigl[ (\partial \phi)^2 +
 (\partial \psi)^2 \Bigr] + \frac12 \Bigl[ m^2 \phi^2 +
 M^2 \psi^2 \Bigr] + \frac{1}{24} \Bigl[ \lambda_\phi \phi^4
 + 6 \,g^2 \phi^2 \psi^2 + \lambda_\psi \psi^4 \Bigr] \,,
\ee
where $M^2 \gg m^2 > 0$.

Since $\psi$ is heavy, we may integrate it out to obtain an
effective theory describing the self-interactions of $\phi$ at
energies below $M$. One of the contributing graphs is given in
Fig.~1, which when evaluated at zero momentum in dimensional
regularization\footnote{We deliberately do not phrase this
discussion in terms of a momentum cutoff, $\Lambda$, as is often
done, for reasons described in more detail elsewhere
\cite{cutoffs}. Rather than meaning that naturalness arguments are
irrelevant, it means as this section shows, that they can be
usefully recast in terms of ratios of renormalized mass
parameters.} leads to
\ba \label{quartic4Destimate}
 \delta m^2 &\simeq& g^2 \mu^{2\epsilon}
 \int \frac{\exd^n k}{(2\pi)^n}
 \left( \frac{1}{k^2 + M^2} \right) + \cdots \nn\\
 &\simeq& c\, \frac{\mu^{2\epsilon}}{2\epsilon}
 \left( \frac{g M}{4 \pi} \right)^2  + \cdots\,,
\ea
where $2\epsilon = n-4 \to 0$ is the dimensional regulator, whose
$1/\epsilon$ divergence is explicitly displayed by factoring it
out of the dimensionless coefficient $c$. This divergent
contribution is renormalized into the light-scalar mass parameter,
$m^2(\mu)$, along with its other UV-divergent contributions.

The renormalized parameter $m^2$ acquires a renormalization-group
(RG) running that at one loop is of order
\be
 \mu \frac{ \partial m^2}{\partial \mu} \simeq
 \Bigl[ c_1 \, g^2 M^2
 + (c_2 \lambda_\phi + c_3 \, g^2) m^2 \Bigr] \,,
\ee
where $c_i$ are dimensionless constants (into which factors of
$1/(4\pi)^2$ are absorbed), and the $m^2$ term arises from the
graph of Fig.~2 using the quartic $\phi^4$ interaction, and from
the (wave-function) renormalization of the kinetic terms. This may
be integrated to give
\ba
 m^2(t) &=& e^{\int^t_0 d x \, (c_2 \lambda_\phi + c_3 g^2)}
 \left[ m_0^2 + c_1 \int^t_0 \exd x \; g^2 M^2 e^{-
 \int^x_0 d u \, (c_2 \lambda_\phi + c_3 g^2)} \right] \nn\\
 &\simeq& m_0^2 \left( \frac{\mu}{\mu_0}
 \right)^{c_2 \lambda_\phi + c_3 g^2} - \frac{c_1 g^2 M^2}{
 c_2 \lambda_\phi + c_3 g^2} \left[ 1 - \left( \frac{\mu}{\mu_0}
 \right)^{c_2 \lambda_\phi + c_3 g^2} \right] \,,
\ea
where $m_0^2 = m^2(\mu = \mu_0)$ and $t = \ln(\mu/\mu_0)$. Here
the second, approximate, equality neglects the $\mu$ dependence of
$\lambda_\phi$, $g^2$ and $M^2$.

This expression shows that $m^2(\mu)$ can only be much smaller
than $M^2$ at scales $\mu \ll M$ if $m^2(M)$ is large -- of order
$g M^2/(4\pi)^2$ -- in order to very precisely cancel with the
evolution from $\mu = M$ to $\mu \ll M$. A parameter like this,
whose small size cannot be understood at any scale $\mu$ where one
chooses to ask the question, is called `technically
unnatural.'\footnote{Just how repulsive this is depends on how
detailed the cancellation must be. Although reasonable people can
differ on how repelled they are by cancellations of 1 part in 100
or 1000, most would agree that cancellations of 5 decimal places
or more would be unprecedented.} Although we know of many
hierarchies of mass in Nature, we know of none for which the small
scale emerges in this kind of technically unnatural
way.\footnote{A possible exception is the 4D cosmological
constant, for which no completely convincing technically natural
proposal has been made. We regard the jury to be out on whether a
technically natural solution to this particular problem is
possible.} Instead, either $m^2$ really is measured to be of order
$g M^2/(4\pi)^2$ (and so is not unnaturally small), or there
exists a (possibly approximate) symmetry \cite{TN} that ensures
that corrections to the parameter $m^2$ are never larger than of
order $m^2$ itself.

The goal of this and later sections is to estimate the size of
these loop corrections to scalar masses from several sources in
extra-dimensional models, viewing them as lower bounds on the
masses of the physical scalars that can naturally emerge from such
models.

\subsubsection*{Relevant and irrelevant interactions}

In the previous example the coupling between light and heavy
sectors was through a marginal interaction, described by the
dimensionless coupling $g$. But there are also other kinds of
dangerous interactions that can generate large corrections to
small scalar masses. One class of these consists of
super-renormalizable -- or relevant, in the RG sense --
interactions, for which the corresponding couplings have dimension
of a positive power of mass (in units where $\hbar = c = 1$).

\centerline{
\begin{picture}(170,70)
    \thicklines
    \put(80,30){\circle{40}}
    \put(60,30){\circle*{4}}
    \put(100,30){\circle*{4}}
    \multiput(57,30)(-7,0){5}{\line(1,0){4}}
    \multiput(100,30)(7,0){5}{\line(1,0){4}}
    \put(21,27){$\phi$}
    \put(135,27){$\phi$}
    \put(78,17){$\psi$}
\end{picture}}
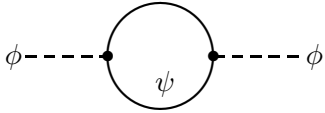
{\bf Figure 2}: A graph contributing to a scalar mass through
cubic couplings. \vspace{0.2cm}

For instance, if one were to supplement the above theory with
super-renormalizable cubic interactions, that break the symmetry
$\phi \to - \phi$,
\be
 -\Delta \L = \frac16 \Bigl[ \xi_\phi \phi^3 + 3h \,\phi\, \psi^2
 \Bigr] \,,
\ee
then evaluating the graph of Fig. 2 gives the new contribution
\be \label{cubic4Destimate}
 \delta m^2 \simeq h^2 \int \frac{\exd^n k}{(2\pi)^n}
 \left( \frac{1}{k^2 + M^2} \right)^2 + \cdots \,,
\ee
and so
\be
 \mu  \frac{\partial m^2}{\partial \mu^2}
  \simeq c' \left( \frac{h^2}{16 \pi^2} \right)  + \cdots\,.
\ee
Barring cancellations one expects $m^2$ to be larger than either
$(h/4\pi)^2$ or $(g M/4\pi)^2$, whichever is largest.

In general the underlying theory could also include
non-renormalizable -- or irrelevant, in the RG sense -- effective
interactions, such as
\be
 -\Delta \L_{\rm nr} = \frac{1}{48\Lambda^2} \, \phi^2
 \, \psi^4 \,,
\ee
where $\Lambda \gg M \gg m$ is some still-higher scale that is
already integrated out to obtain our starting lagrangian,
eq.~\pref{phipsiL}. This kind of interaction generates a
contribution to the light scalar mass (see Fig. 3) that is of
order
\be
 \delta m^2 \simeq \frac{1}{\Lambda^2} \left[ \int
 \frac{\exd^n k}{(2\pi)^n} \; \frac{1}{k^2 + M^2}
 \right]^2 \propto \left( \frac{1}{16\pi^2} \right)^2
 \frac{M^4}{\Lambda^2} \,,
\ee
and so contributes contributions to $m^2$ that are small relative
to those already considered.

\centerline{
\begin{picture}(170,70)
    \thicklines
    \put(80,30){\circle{40}}
    \put(80,10){\circle*{4}}
    \put(80,-10){\circle{40}}
    \multiput(77,10)(-9,0){5}{\line(1,0){4}}
    \multiput(80,10)(9,0){5}{\line(1,0){4}}
    \put(21,7){$\phi$}
    \put(135,7){$\phi$}
    \put(78,21){$\psi$}
    \put(78,-5){$\psi$}
\end{picture}}
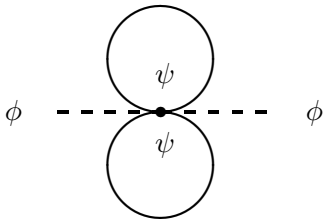
\vspace{1cm} {\bf Figure 3}: A large mass correction to a light
scalar from an irrelevant coupling.

\subsection{Extra dimensions without SUSY}

Two new issues that arise when considering naturalness in
extra-dimensional models (see refs.~\cite{EDLoops} for other
discussions of loops in extra dimensions) are higher-dimensional
kinematics and symmetries. An extra-dimensional generalization of
the two-scalar model described above provides the simplest context
for discussing the first of these, while higher dimensional
gravity provides the most commonly encountered framework for the
second. We therefore consider each of these examples in turn.

\subsubsection*{Scalar fields}

Start first with a light and heavy scalar field in $D = 4 + d$
dimensions, with lagrangian
\be
 -\L = \frac12 \Bigl[ (\partial \Phi)^2 + (\partial \Psi)^2
 \Bigr] + \frac12 \Bigl[ m^2 \Phi^2 + M^2 \Psi^2 \Bigr]
 + \frac{g_3}{2} \, \Phi \Psi^2 + \frac{g_4^2}{4} \,
 \Phi^2 \Psi^2 + \cdots \,,
\ee
and use (for now) a flat background metric
\be
 \exd s^2 = \eta_{\mu\nu} \, \exd x^\mu \exd x^\nu
 + L^2 \delta_{mn} \, \exd y^m \exd y^n \,,
\ee
whose extra-dimensional volume is $V = L^d$. Since a canonically
normalized scalar field in $D$ dimensions has engineering
dimension (mass)${}^{(D-2)/2}$, the cubic and quartic couplings
generically are RG-irrelevant, having negative mass dimension
\be
 g_3 \sim \left( \frac{1}{\Lambda} \right)^{(D-6)/2}
 \quad \hbox{and}\quad
 g_4 \sim \left( \frac{1}{\Lambda} \right)^{(D-4)/2} \,.
\ee

We next integrate out the massive field $\Psi$, assuming the
masses satisfy the hierarchy\footnote{We do not attempt to track
factors of $2\pi$, so ignore these in the definition of $M_\KK$.}
$M^2 \gg m^2 \gg M_\KK^2 := 1/L^2$. In this limit the sum over
discrete KK modes is well described by a continuous integral, and
so the contribution of the cubic interaction (from the graph of
Fig.~2) is of order
\be
 \delta m^2 \simeq g_3^2 \int \frac{\exd^n k}{(2\pi)^n}
 \left( \frac{1}{k^2 + M^2} \right)^2 \propto  g_3^2 M^{D-4}
 \simeq \left( \frac{M}{\Lambda} \right)^{D-6} M^2 \,,
\ee
where this time $n = D - 2\epsilon \to D$ rather than 4. The
result obtained from using the quartic interaction in Fig.~1 is
similarly estimated to be
\be
 \delta m^2 \simeq g_4^2 \int \frac{\exd^n k}{(2\pi)^n}
 \left( \frac{1}{k^2 + M^2} \right) \propto g_4^2 M^{D-2}
 \simeq \left( \frac{M}{\Lambda} \right)^{D-4} M^2 \,.
\ee
These both differ by the factor $(M/\Lambda)^{D-4} =
(M/\Lambda)^d$ relative to the corresponding 4D example (for which
$d= 0$), and these extra powers of $M$ arise ultimately due to the
additional phase space available for the extra-dimensional loop
momenta.\footnote{Alternatively these additional powers of $M$
relative to the 4D results found earlier can be regarded as being
due to the necessity to sum over the tower of KK modes (see
Appendix A).} If $M \simeq \Lambda$ then the corrections are
generically of order $M^2$ in any dimension, but a proper
exploration of contributions at $\Lambda$ should really be done
using whatever UV completion kicks in at this scale. The main
lesson here is that extra-dimensional kinematics make most
interactions irrelevant in the technical sense, leading to higher
powers of the heavy scale $M$ in the generic contribution to
$\delta m^2$.

\subsubsection*{Gravity}

In practice, the fields of most interest in higher dimensions are
various types of gauge and gravitational fields. The crucial
difference between these and the scalar fields described
heretofore is the existence of local gauge symmetries (or general
covariance) that keep these fields massless in the
extra-dimensional theory. Their only masses in 4D are therefore
those due to the KK reduction. We now explore the sensitivity of
these masses to integrating out particles that are both heavier
and lighter than the KK scale.

To see the effects of integrating out a heavy field in this case
consider a massive scalar coupled to gravity,
\be
 - \frac{\L}{\sqrt{-g_{(\ssD)}}} = \lambda_0 + \frac{1}{2\kappa_0^2} \, \R
 + \frac12 (\partial \Psi)^2 + M^2 \Psi^2 \,,
\ee
where $\R = g^{\ssM \ssN} \R_{\ssM \ssN}$ denotes the metric's
Ricci scalar, $\lambda_0$ is a (bare) cosmological constant, and
$\kappa_0^2 = 8\pi G_\ssD := M_g^{-(D-2)/2}$ is the (bare) reduced
gravitational coupling constant, which also defines the
higher-dimensional Planck scale, $M_g$. A semiclassical treatment
\cite{PCGrav} assumes we require $\lambda \ll M_g^D$ and that the
heavy particle mass satisfies $M \ll M_g$.

As before, integrating out the massive field $\Psi$ leads to many
new effective interactions for the remaining light field, which
are local so long as the heavy particle's Compton wavelength, is
much shorter than the size of the extra dimensions, $M \gg M_\KK$.
These can be organized in order of increasing dimension, with the
coefficient of each involving the appropriate power of $M$, as
required on dimensional grounds. Because in the present case the
low-energy field is the metric, the form of these interactions is
strongly restricted by general covariance to take the form of a
curvature expansion,
\be \label{curvatureexpn}
 - \frac{\L_{\rm eff}}{\sqrt{-g_{(\ssD)}}} = \lambda
 + \frac{1}{2\kappa^2} \, \R
 + a_2 M^{D-4} \R_{\ssM\ssN\ssP\ssR}
 \R^{\ssM\ssN\ssP\ssR} + \cdots \,,
\ee
where all possible terms involving curvatures and their
derivatives are contained in the ellipses, while $\lambda$ and
$\kappa^2$ are appropriately renormalized
\be
 \lambda = \lambda_0 + a_0 M^D
 \quad \hbox{and} \quad
 \frac{1}{2\kappa^2} = \frac{1}{2\kappa_0^2} + a_1 M^{D-2} \,,
\ee
with $a_i$ dimensionless constants, and so on.

In general, the condition $M \ll M_g$ implies $\kappa^2 \simeq
\kappa^2_0$, making the Einstein term largely insensitive to the
effects of integrating out heavy particles. The same is usually
not true for $\lambda$ and $\lambda_0$ because the Einstein
equations imply $\R \simeq \O(\kappa^2 \lambda)$, and
calculability demands $\R \simeq 1/L^2 = M_\KK^2$ be much smaller
than $M_g^2$. For non-supersymmetric theories taking $M \gg 1/L$
and $\kappa^2 \lambda \simeq \O(M_\KK^2)$ usually means
fine-tuning $\lambda_0$ to ensure that $\lambda \sim \O(M_\KK^2
M_g^{D-2}) \ll M^D$.

The upshot is this: integrating out a field with mass $M \gg
M_\KK$ just leads to the addition of local interactions to the
higher-dimensional action. These do not qualitatively change the
low-energy consequences so long as it is the lowest-dimension
interactions that are of physical interest (like the Einstein
action) and provided that the most general interactions are
included in the action from the start. This is because the
condition for the validity of the semiclassical treatment in terms
of an extra-dimensional field theory requires $M \ll M_g$,
ensuring that the new contributions are swamped by those involving
$M_g$. An understanding of the contributions at scale $M_g$ then
should be done using the UV completion at the gravity scale, which
within string theory would potentially involve a full string
calculation.

\subsubsection*{Integrating out KK modes}

The effective theory becomes four-dimensional at energies of order
$M_\KK$ and below, so once these scales are integrated out we can
no longer use higher-dimensional symmetries (like
extra-dimensional general covariance) to restrict the form of the
result.

We wish to track how the integrating out of modes with masses at
the KK scale and below depends on the underlying scales $M_\KK$ or
$M_g$. Within the semiclassical approximation any such an
integration arises as an expansion about a background field,
$\overline{g}_{\ssM\ssN}$, so that
\be
 g_{\ssM\ssN} = \overline{g}_{\ssM\ssN} + \kappa h_{\ssM \ssN} \,.
\ee
It is also useful to explicitly scale out the local linear size of
the extra dimensions, $e^{u(x)}$, (measured using the background
geometry) from the total metric
\be
 g_{\ssM\ssN} \exd x^\ssM \exd x^\ssN
 = \omega \, e^{-du} \, \hat g_{\mu\nu} \exd
 x^\mu \exd x^\nu + e^{2u} \hat g_{mn} \exd y^m \exd y^n +
 \hbox{off-diagonal terms} \,,
\ee
where the factor pre-multiplying the 4D metric is chosen to ensure
no $u$-dependence in the 4D Einstein action ({\em i.e.} the 4D
Einstein frame).

The factor $\omega := (M_p/M_g)^2$ numerically converts to 4D
Planck units, and the vacuum value $\langle e^u \rangle \propto
M_g L$ provides a dimensionless measure of the extra-dimensional
linear size\footnote{We take $V \simeq L^d$ and $R \simeq 1/L^2$
for the same scale $L$, and by so doing ignore features that might
arise within strongly warped spacetimes (for which the low-energy
sector can be strongly localized), or within some negatively
curved spaces (for which the scales associated with volume and
curvature can radically differ).} (and so $\langle e^{du} \rangle
\simeq (M_g L)^d := \V$). Recall the 4D Planck scale is related to
the dimensionless volume, $\V = V M_g^{d}$, by $M_p^2 = V
M_g^{D-2} = \V M_g^2$, and so $\omega = {M_p^2}/{M_g^2} = {\V}$,
and $M_g \simeq M_p/\V^{1/2}$. Using $\sqrt{- g_{(\ssD)}} =
\sqrt{- \hat g_{(4)}} \, \sqrt{\, \hat g_{(d)}} \; \omega^2
e^{-du}$ and $\int \exd^d y \propto M_g^{-d}$, we find the 4D
Einstein term becomes
\ba
 M_g^{D-2} \int \exd^d y \sqrt{- g_{(\ssD)}} \;
 g^{\mu\nu} \R_{\mu\nu} &=& \omega
 \, M_g^2 \sqrt{- \hat g_{(4)}} \;
 \hat g^{\mu\nu} \hat R_{\mu\nu} + \cdots \nn\\
 &=& M_p^2 \sqrt{- \hat g_{(4)}} \; \hat g^{\mu\nu}
 \hat R_{\mu\nu} + \cdots \,,
\ea
as required.

Expanding the action in powers of fluctuations and focussing on
the 4D scalar KK modes contained within ${h^m}_p :=
\overline{g}^{mn} h_{np}$ --- generically denoted $\varphi^i$ ---
similarly gives the following dimensionally reduced kinetic terms
\ba
 -\L_{\rm kin} \simeq \frac12 M_g^{D-2} \int \exd^d y
 \sqrt{- g_{(\ssD)}} \; g^{\mu\nu} \R_{\mu\nu}
 &=& \frac\omega2 \, M_g^2 \int \exd^d y
 \sqrt{- \hat g_{(4)}} \; \hat g^{\mu\nu} {\H^{mn}}_{pq} \
 \partial_\mu {h^p}_m \partial_\nu {h^q}_n + \cdots \nn\\
 &:=& \frac{M_p^2}{2} \sqrt{- \hat g_{(4)}} \;
 \hat g^{\mu\nu} \G_{ij}(\varphi)
 \partial_\mu \varphi^i \partial_\nu \varphi^j + \cdots \,.
\ea
where ${\H^{mn}}_{pq}$ are a set of coefficients depending on the
${h^m}_p$ (but not their derivatives), while $\G_{ij}(\varphi)$
denotes the target-space metric for the dimensionless 4D fields
$\varphi^i$. The detailed form of $\G_{ij}$ is not important
beyond the fact that it contains no additional dependence on $\V$,
and so is generically $\O(1)$ in the large-$\V$ limit.

By contrast, the contributions to the scalar potential for the
$\varphi^i$ instead scale with $M_g$ and $L$ according to
\ba
 &&-\L_{\rm pot} \simeq M_g^{D-2}
 \int \exd^d y \sqrt{- g_{(\ssD)}} \;
 g^{mn} \R_{mn} \nn\\
 && \qquad\qquad\qquad = \omega^2 \,
 M_g^2 \int \exd^d y \, e^{-du} \sqrt{- \hat g_{(4)}} \;
 (e^{-2u} \hat g^{mn}) {\K^{pq}}_{rs} \,
 \partial_m {h^r}_p \,\partial_n {h^s}_q + \cdots \nn\\
 && \qquad\qquad\qquad :=
 \frac{M_p^4}{\V^{1+2/d}} \sqrt{- \hat g_{(4)}}
 \; U(\varphi) \nn\\
 && \qquad\qquad\qquad =
 M_\KK^2 M_p^2 \sqrt{- \hat g_{(4)}} \; U(\varphi) \,,
\ea
which uses $e^{(d+2)u} = \V^{1+2/d}$ and $M_\KK^2/M_p^2 =
(M_\KK^2/M_g^2) (M_g^2/M_p^2) \simeq \V^{-2/d} \, \V^{-1}$. Again,
the detailed form of ${\K^{pq}}_{rs}$ and $U(\varphi)$ are not
important, beyond the fact that they generically do not contribute
to the $\V$ dependence of the result.

Once canonically normalized, $\varphi \propto \phi /M_p$, we find
the following schematic mass terms, cubic and quartic interactions
\ba \label{KKKouplings}
 - \frac{\L_2}{\sqrt{-g_{(4)}}} &\simeq& (\partial \phi)^2 +
 M_\KK^2 \, \phi^2 \nn\\
 -\frac{\L_3}{\sqrt{-g_{(4)}}} &\simeq&  \frac{1}{M_p}
 \phi (\partial \phi)^2
 + \frac{M_\KK^2}{M_p} \, \phi^3 \\
 \hbox{and} \quad
 - \frac{\L_4}{\sqrt{-g_{(4)}}} &\simeq& \frac{1}{M_p^2}
 \phi^2 (\partial \phi)^2 + \frac{M_\KK^2}{M_p^2} \, \phi^4
 \,, \nn
\ea
and so on. These show that the generic mass is $M_\KK \simeq
M_p/\V^{(1+2/d)/2}$ (as expected), and although the low-energy
{\em derivative} interactions are Planck suppressed, those in the
scalar potential have a universal additional suppression by a
factor of $M_\KK^2/M_p^2 = 1/\V^{1+2/d} = (M_g^2/M_p^2)^{1 + 2/d}$
relative to generic Planck size.

A similar analysis for the curvature-squared terms shows that
these introduce three kinds of 4D interactions: ($i$) $\O(1)$
4-derivative interactions, $\sim k_4(\varphi) (\partial
\varphi)^4$; ($ii$) $\O(M_\KK^2)$ two-derivative interactions,
$\sim M_\KK^2 k_2(\varphi) (\partial \varphi)^2$; and
$\O(M_\KK^4)$ potential terms, $\sim M_\KK^4 k_0(\varphi)$. Each
is therefore suppressed relative to its counterpart (if this
exists) coming from the Einstein term by an additional factor of
$M_\KK^2/M_p^2 = 1/\V^{1+2/d}$.

Provided $\lambda$ is chosen to allow classical solutions for
which $\R \simeq \kappa^2 \lambda \simeq 1/L^2 \sim M_\KK^2$, as
discussed above, the contributions of the cosmological constant
term to the 4D dimensionally reduced interactions scale in the
same way as do those coming from the Einstein term.

\subsubsection*{Naturality of the KK couplings}

It is noteworthy that, from a 4D perspective, the couplings of the
KK modes amongst themselves are {\em weaker} than Planckian, yet
they are stable against radiative corrections.

As we've seen, the dominant contribution from the scales much
larger than $M_\KK$ come from scales $M \gsim M_g$, which must be
performed within the theory's UV completion. If this is a string
theory, the contribution to the low-energy action from integrating
out string states is what gives the initial higher-dimensional
gravity action, plus a variety of higher-derivative corrections
involving powers of the curvature and other low-energy fields.
Since our initial estimate for the size of the interactions comes
from the higher-dimensional Einstein term, the leading corrections
in the UV come from dimensionally reducing terms involving more
derivatives than this. If it is a curvature-squared term that
dominates, then we expect the largest corrections from these
scales to be suppressed relative to the leading terms by of order
$M_\KK^2/ M_g^2$.

The sole exception to this happy picture of insensitivity to
higher-scale physics is the contribution to the extra-dimensional
cosmological constant, although this is also not dangerous once
$\lambda \simeq M_\KK^2/\kappa^2 \simeq M_\KK^2 M_g^{D-2}$ is
tuned to be small enough to allow the extra dimensions to be large
in the first place. As we shall see, even this problem need not
arise when couched in a supersymmetric context, since in this case
higher-dimensional supersymmetry usually requires $\lambda = 0$.

Finally, what of the naturality of the 4D scalar masses from loops
with $M \lsim M_\KK$? Since these are not protected by
higher-dimensional symmetries, they should be analyzed within the
effective 4D theory. As shown above, it is the contributions of
the relevant and marginal -- {\em i.e.} cubic and quartic --
interactions that are then the most dangerous. But because all of
the interactions in the scalar potential are suppressed by
$M_\KK^2 /M_p^2$ relative to Planck strength, their use in loop
graphs generates interactions that are generically suppressed
relative to those we start with by similar factors. This implies
they are of similar order to the contributions of
higher-derivative corrections in the extra-dimensional theory just
considered.

Similarly, graphs using the derivative interactions give much the
same result, since although these involve couplings unsuppressed
(relative to $M_p$) by powers of $1/\V$, they are more UV
divergent and so depend more strongly on the mass of the heaviest
4D state that can circulate within the loop. But this mass is
again $M_\KK$, since for masses much higher than this the
restrictions of higher-dimensional general covariance limit the
result to one of the local interactions considered above. For
instance, using these estimates in Fig.~2 then gives
\be
 \delta m^2 \simeq \frac{1}{M_p^2}
 \int \frac{\exd^4k}{(2\pi)^4} \; \frac{k^4}{(k^2 + m^2)^2}
 \simeq \frac{c \mu^{2\epsilon}}{\epsilon}
 \, \left( \frac{M_\KK^4}{M_p^2} \right) \,,
\ee
again implying corrections that are suppressed by $M_\KK^2/M_p^2$.

These arguments indicate that the dominant corrections to the 4D
scalar potential are of order $\delta V \simeq \O(M_\KK^4) \delta
U(\varphi)$. Notice that this implies the corresponding
contributions to the 4D vacuum energy are $\delta V \simeq M_\KK^4
\simeq 1/L^4$, in agreement with explicit Casimir energy
calculations \cite{Casimir}.

\subsubsection*{Moduli}

It often happens that accidental symmetries in the leading order
action imply that some of the KK scalars are massless once
dimensionally reduced. This would happen for the volume modulus,
$e^u$, in particular, if the equations of motion coming from the
leading action were scale invariant (as would be true for the
Einstein equations in the absence of a cosmological term, or for
the lowest-derivative terms in most higher-dimensional
supergravities).

In this case the dimensionally reduced mass for the corresponding
moduli comes from next-to-leading effects that break the relevant
symmetry. These could be from loop effects in the low-energy 4D
theory, in which case the above estimates indicate their masses
would be expected to be of order $M_{\rm mod} \simeq M_\KK^2/M_p
\ll M_\KK$ rather than being precisely massless.\footnote{Notice
that if $M_\KK \simeq 10^{-3}$ eV then $M_{\rm mod} \lsim
10^{-30}$ eV, showing that if the observed vacuum energy density
could be arranged to be dominated by a KK energy, then the
presence of moduli would produce a quintessence cosmology, with no
additional tunings required to ensure small enough scalar masses
\cite{NatQuin}.} Alternatively, the dominant corrections could
come from loop effects in the higher-dimensional theory, which
we've seen are equivalent to use of sub-dominant terms in the
derivative expansion (such as from curvature-squared or higher)
when compactifying. If it is curvature-squared terms that
dominate,\footnote{For an example where curvature-squared terms do
{\em not} dominate, see the supersymmetric estimate below.} then
the discussion above shows we again expect masses of order $M_{\rm
mod} \simeq M_\KK^2/M_p$. Notice that masses these size are also
no larger than the generic size of radiative corrections in the
low-energy theory, based on the estimates given above.

\subsection{Supersymmetric effects}

For supersymmetric systems, the estimates just given can differ in
several important ways.

\begin{itemize}
\item {\it Extra-dimensional non-renormalization theorems:} For
higher dimensional supergravity it is the expectation of one of
the fields (the dilaton) that plays the role of the loop-counting
parameter. Often its appearance in the leading derivative
expansion of the action is restricted by supersymmetry, in which
case extra-dimensional loops must necessarily contribute
suppressed both by powers of the coupling constant {\em and}
low-energy factors. For instance, in string theory it is often
true that higher-order contributions in the string coupling,
$g_s$, necessarily only arise for terms that are also sub-leading
in powers of the string scale, $\alpha' = 1/M_s^2$
\cite{LoopAlpha'}.

It is also true that supersymmetry can raise the order in the
derivative expansion of the first subdominant contributions to the
action that arise even without loops. For instance, in the Type
IIB models of later interest, the first higher-curvature
corrections that arise at string tree level involve four powers of
the curvature, as opposed to the generic expectation of curvature
squared \cite{TreeAlpha'}.
\item {\it Four-dimensional non-renormalization theorems:} If the
supersymmetry breaking scale should be much smaller than $M_\KK$,
then the effective 4D description can be written as a 4D
supergravity, possibly supplemented by soft supersymmetry-breaking
terms. In this case the most dangerous relevant and marginal
scalar interactions appear in the holomorphic superpotential,
$W(\phi)$, and so are protected by 4D non-renormalization theorems
\cite{NRT1,NRT2,NRT3,NRT4}. If supersymmetry is not broken these
exclude corrections to all orders in perturbation theory.
Corrections become possible once supersymmetry is broken, but are
further suppressed by the relevant supersymmetry breaking scale.
When computing explicit loops, these suppressions come about
through the usual cancellations of bosons against fermions,
together with mass sum rules like \cite{sumrule}
\be \label{sumrule}
 \sum_s (-)^{2s} (2s+1) \hbox{Tr } M_s^2 \simeq M_{3/2}^2 \,,
\ee
where $M_{3/2}$ is the 4D gravitino mass.
\item {\it Additional extra-dimensional fields:} A third way in
which supersymmetric theories can differ is through their
extra-dimensional field content, which always involves more fields
than just gravity. Interactions involving these other fields
sometimes provide the dominant contributions to quantities like
the masses of moduli. The most familiar example of this sort
occurs when the background supergravity solution involves nonzero
3-form flux fields, $G_{mnp}$, whose presence gives some of the
moduli masses (as in 10D Type IIB flux compactifications). In this
case it is the term $\L = \sqrt{-g} \; G^{mnp} G_{mnp} \propto
1/\V^2$ that dominates these masses, leading to $M_{\rm mod}
\simeq M_p/\V \gg M_\KK^2/M_p \simeq M_p/\V^{4/3}$.
%
\end{itemize}

\section{Large-volume (LV) string models}

The large-volume (LV) framework {\cite{LV}} is a scenario for
moduli stabilization for Calabi-Yau compactifications within Type
IIB string theory. In generic Type IIB flux compactifications, the
presence of background 3-form fluxes alone can stabilize the
dilaton and the complex structure moduli of the underlying
Calabi-Yau geometry \cite{GKP}. The K\"ahler moduli can then be
stabilized by the non-perturbative effects localized on four
cycles associated with various branes that source the geometries
\cite{KKLT}. The LV scenario identifies an interesting subclass
for which the volume modulus is naturally stabilized at
exponentially large volumes,
\be
 \V \propto \exp\left( c \tau_s \right)  \,,
\ee
where $\tau_s$ is the size of a (comparatively much smaller)
blow-up cycle of a point-like singularity in the underlying
geometry, while $c$ is an order-unity constant.

The framework applies to a large class of Calabi-Yau
compactifications since there are only two requirements for its
implementation \cite{CCQTWO}: ($i$) there must be at least one of
the K\"ahler moduli must be the blow-up mode, $\tau_s$, of a
point-like singularity; and ($ii$) the number of complex structure
moduli have to be greater than the the number of K\"ahler moduli
(in order for the Euler number of the Calabi-Yau space to have the
sign required for the potential to have a minimum for large $\V$).

A key ingredient is the inclusion of the leading order $\alpha'$
correction to the 4D K\"ahler potential, since this is what
generates the potential with a minimum with compactification
volumes that are exponentially large in $\tau_s$. Furthermore,
because $\tau_s$ scales as the inverse of the the value of the
dilaton,
\be
 \tau_s \propto \frac{1}{g_s} \,,
\ee
and so is given as a ratio of integer flux quanta. Thus the
framework naturally generates an exponentially large volume of
compactification from integer flux quanta, with $\V$ passing
through an enormous range of values as the fluxes range through a
range of order tens.

In what follows we review some relevant aspects of these
compactifications with an emphasis on the mass scales and strength
of couplings.

\subsection{Masses and couplings}

LV models enjoy a rich pattern of particle masses and couplings,
for which it is useful for many purposes to express in 4D Planck
units. Our goal here is to express how these quantities scale as
functions of the dimensionless extra-dimensional volume, expressed
in string units: $\V \simeq \alpha'^3 \int \sqrt{g_6} \simeq
(L/\ell_s)^6 \gg 1$.

\subsubsection*{Masses}

In terms of $\V$ and the 4D Planck scale, $M_p$, the largest scale
in the excitation spectrum is the string scale itself, $M_s^{-2} =
\ell_s^2 \simeq \alpha'$, which (ignoring factors of $g_s$) is of
order
\be
 M_s \simeq \frac{M_p}{\V^{1/2}} \,.
\ee
This characterizes the order of magnitude of the masses of generic
string excitation levels.

If the linear sizes of the various extra dimensions are all of the
same order, $L$, then since $\V \propto (L/\ell_s)^6$, the scaling
with $\V$ of the masses of Kaluza-Klein (KK) excitations is
\be
 M_\KK \simeq \frac 1L \simeq \frac{M_s}{\V^{1/6}}
 \simeq \frac{M_p}{\V^{2/3}} \,.
\ee

The presence of fluxes stabilizes the complex structure moduli, or
what would otherwise have been massless KK zero modes. These are
systematically light compared with the KK scale because of the
$\V$ suppression of the various fluxes, due to the quantization
conditions of the form $\int F \simeq N \alpha'$, for an
appropriate 3-form, $F_{mnp}$. Masses obtained from the flux
energy, $\int {\rm d}^6x \sqrt{-g_{10}} \; F^2 \propto \V^{-2}$,
are of order
\be
 M_{cs} \simeq \frac{M_s}{\V^{1/2}} \simeq
 \frac{M_p }{\V}\,.
\ee

The K\"ahler moduli survive flux compactification unstabilized to
leading order, and so naively might be expected to receive a mass
that is parametrically smaller than those of the complex-structure
moduli. For instance, small K\"ahler moduli like $\tau_s$ are
stabilized by the nonperturbative terms, $W \simeq e^{-c \tau_s}$,
in the low-energy 4D superpotential, and at the potential's
minimum these are comparable to the $\alpha'$ corrections,
implying the relevant part of the potential depends on the volume
like $1/\V^3$. However, because the K\"ahler potential for these
moduli is $K = -2 \ln \V$, their kinetic term is also proportional
to $1/\V$, implying that the volume-dependence of the mass of
these moduli is again of order
\be
 M_k \simeq \frac{M_s}{\V^{1/2}} \simeq
 \frac{ M_p}{\V} \,.
\ee

The K\"ahler modulus corresponding to the volume of the
compactification also receives its mass from the leading $\alpha'$
correction. However for this modulus $\partial \V$ depends
nontrivially on $\V$ (unlike for the small moduli like $\tau_s$),
leading to a kinetic term of order $\partial \overline \partial K
\simeq \partial \overline
\partial \V/\V$ and a quadratic term in the potential that
is of order $\partial \overline \partial \V/ \V^4$. The result is
a volume-modulus mass whose $\V$-dependence is of order
\be
 M_\ssV \simeq \frac{M_s}{\V} \simeq \frac{M_p}{\V^{3/2}} \,.
\ee

For many compactifications K\"ahler moduli come with a range of
sizes, and there are often many having volumes $\tau_i \gg
\tau_s$. These tend to be exponentially suppressed in the
superpotential, $W \simeq e^{-c_i \tau_i} \ll e^{-c \tau_s} \simeq
1/\V$, and so at face value also appear to have exponentially
small masses when computed using only the leading $\alpha'$
corrections. However for any such moduli it is the sub-leading
corrections to the scalar potential (like string loops or higher
orders in $\alpha'$) that instead dominate their appearance in the
potential. Consequently these states are systematically light
relative to the generic moduli discussed above. The leading
string-loop contribution turns out to be dominant, and so these
moduli generically have masses of order
\be
 M \simeq \frac{M_p}{\V^{5/3}} \,.
\ee

Finally, the gravitino mass associated with supersymmetry breaking
is itself of order
\be
 M_{3/2} \simeq \frac{M_s^2}{M_p}
 \simeq \frac{ M_s}{\V^{1/2}} \simeq
 \frac{M_p}{\V} \,.
\ee
Notice for the purposes of the naturalness arguments that the
masses of the volume modulus and the other large K\"ahler moduli
are parametrically lighter than both the gravitino mass and the KK
scale.

\subsubsection*{Couplings}

The potential term associated with the complex structure and
K\"ahler moduli scales as ${ {\V}^{-2}}$ as
it arises by dimensionally reducing a higher-dimensional flux
energy, $\sim \int \exd^6y \sqrt{g} \; G_{mnp}G^{mnp}$, rather
than from a curvature-squared term.

Because the KK scale is larger than typical SUSY-breaking scales,
like $M_{3/2}$, the effective 4D theory can be described within
the formalism of $N=1$ supergravity. In this context the
suppression of the couplings relative to Planck strength reflects
itself in the no-scale form of the Kahler potential,
\ba
  K = - 2 \log \V \,,
\ea
since this suppresses all terms in the potential by a factor of
$e^{K} W \propto W/\V^2$.

The volume dependence of the interactions of low-energy fields
with Kaluza-Klein modes can be inferred using arguments
\cite{Steve} very similar to those used in section
\ref{sec:GenLoopEst}. Specializing the potential term obtained
there to $D = 10$ and $d = 6$ implies that it scales as
$\V^{-4/3}$.

The couplings of these fields to ordinary matter can be estimated
if the ordinary matter is assumed to be localized on a
space-filling brane somewhere in the extra dimensions. The
strength of these couplings depends on the particular brane field
of interest, but a representative coupling is to gauge bosons,
which has the generic form
\be
 {\cal L}_{\rm int} = \frac{\phi}{f} \,
 F^{\mu\nu} F_{\mu\nu} \,,
\ee
where, for example, $f \simeq M_p$ for many of the K\"aher moduli
(like the volume modulus itself), although some of the `small'
moduli couple with strength $f \simeq M_s \simeq M_p/\V^{1/2}$.

\section{Radiative corrections in LV models}

The previous sections summarize the rich pattern of scalar masses,
related to one another by powers of $\V$, predicted by LV models
at leading order:
\be
 M_s \simeq \frac{M_p}{\V^{1/2}} \,, \quad
 M_\KK \simeq \frac{M_s}{\V^{1/6}} \simeq
 \frac{M_p}{\V^{2/3}} \,, \quad
 M_{\rm mod} \simeq M_{3/2} \simeq \frac{M_p}{\V} \,, \quad
 M_\ssV \simeq \frac{M_p}{\V^{3/2}} \,.
\ee

The couplings among these states are also $\V$-dependent, and this
is important when computing the size of the loop-generated masses.
The contributions of wavelengths longer than the KK scale to these
loop corrections can be evaluated within the context of an
effective 4D theory, while those having shorter wavelength should
be computed within the higher dimensions. Technical naturalness
asks that both the 4D and the higher-dimensional contributions be
smaller than the lowest-order mass of the light particle itself.

\subsection{4D contributions}

Suppose a light particle has a mass, $m_\phi \simeq M_p \V^{-a}$,
that is suppressed by a particular negative power of $\V$. Suppose
further that this light particle couples to a more massive
particle having a mass $m_\psi \simeq M_p \V^{-b}$, with $b < a$.
Imagine these particles to experience a relevant cubic coupling of
the schematic form $\L_3 \simeq h \, \phi \, \psi \partial^{p_3}
\psi$ whose coupling strength is of order $h \simeq M_p
(1/M_p)^{p_3} \, \V^{- c_3}$, where the power of energy, $E$,
would arise if the coupling were to involve the derivatives of the
fields. Assuming 4D kinematics are relevant, repeating the steps
of section 2 shows that the one-loop correction to the light
scalar mass generated using this interaction in a graph like Fig.
2 is of order
\ba
 \delta m_\phi^2 \simeq
 h^2 \int \frac{\exd^4 k}{(2 \pi)^4}
 \; \frac{k^{2 p_3}}{(k^2 + m_\psi^2)^2}
 &\simeq& \left( \frac{h \, m_\psi^{p_3}}{4\pi}
 \right)^2  \nn\\
 &\simeq& \frac{1}{(4\pi)^2}
 \left( \frac{1}{\V} \right)^{2(p_3 b + c_3)} \, M_p^2 \,.
\ea

The mass of the light particle is larger than these radiative
corrections provided $\delta m_\phi \lsim m_\phi$, and so $b \,
p_3 + c_3 \ge a$. In light of the discussion in \S2, since the
most massive states for which 4D kinematics applies have $m_\psi
\simeq M_\KK \simeq M_p/\V^{2/3}$, we make take as the worst case
$b = \frac23$. Furthermore, the discussion of section
\ref{sec:GenLoopEst} implies that the couplings of such states is
generically either a Planck-strength derivative coupling ($p_3 =
2$ and $c_3 = 0$), or a potential interaction whose strength is
proportional to $M_\KK^2/M_p^2 \simeq \V^{-4/3}$ ($p_3 = 0$ and
$c_3 = \frac43$). In both cases $b \, p_3 + c_3 = \frac43$,
implying corrections that are sufficiently small for all 4D moduli
(for which $a = 1$) except for the volume modulus (for which $a =
\frac32$).

Alternatively, suppose the light particle couples to the massive
one through a marginal quartic coupling $\L_4 \simeq g^2 \phi^2
\psi \partial^{2p_4} \psi$, whose coupling strength is of order $g
\simeq (1/M_p)^{p_4} \, \V^{-c_4}$, where the power of energy,
$E$, again arises as derivatives of the fields. In this case the
one-loop correction (using 4D kinematics) to the light scalar mass
is of order
\ba
 \delta m_\phi^2 \simeq
 g^2 \int \frac{\exd^4 k}{(2 \pi)^4}
 \; \frac{k^{2p_4}}{(k^2 + m_\psi^2)} &\simeq&
 \left( \frac{g \, m_\psi^{p_4+1}}{4\pi} \right)^2 \nn\\
 &\simeq& \frac{1}{(4\pi)^2}
 \left( \frac{1}{\V} \right)^{2(p_4+1)b+2c_4} \, M_p^2 \,,
\ea
so this contribution to the mass is technically natural provided
$b \left(1 + p_4 \right) + c_4 \ge a$.

Again, the maximum mass appropriate to 4D kinematics is $m_\psi
\sim M_\KK$, corresponding to $b = \frac23$ and there are two
kinds of quartic interactions amongst the KK modes:
Planck-suppressed derivative couplings ($p_4 = 1$ and $c_4 = 0$)
and non-derivative interactions suppressed by $M_\KK^2/M_p^2$
($p_4 = 0$ and $c_4 = \frac23$). Both of these choices satisfy $b
\left( 1 + p_4 \right) + c_4 = \frac43$, and so are the same size
as those obtained from cubic interactions, and not larger than any
of the moduli masses ($a = 1$), except for the volume modulus ($a
= \frac32$).

\subsubsection*{Supersymmetric cancellations}

Apart from the volume modulus, we see that moduli masses tend to
be generically stable against 4D radiative corrections. Naively
this would seem to indicate the classical approximation is not a
good one for the volume modulus itself, which should then be
expected to be more massive than $\O(M_p/\V^{3/2})$. However to
this point we have not yet availed ourselves of the cancellations
implied by 4D supersymmetry, which survives into the 4D theory
because $M_{3/2} \ll M_\KK$.

Let us reconsider the 4D contributions in this light. As usual,
since particle masses are driven by the form of the
superpotential, they receive no corrections until supersymmetry
breaks. At one loop an estimate of the leading
supersymmetry-breaking effects obtained by summing the
contributions of bosons and fermions, keeping track of their mass
difference. For instance, keeping in mind the sum rule,
eq.~\pref{sumrule}, gives
\be
 \delta m^2 \simeq g \left( m_\ssB^2 - m_\ssF^2 \right)
 \lsim \left( \frac{M_\KK^2}{M_p^2} \right) M_{3/2}^2
 \simeq \left( \frac{1}{\V^{4/3}} \right)
 \frac{M_p^2}{\V^2} \,,
\ee
implying the generic supersymmetry-breaking mass shift is $\delta
m \simeq M_p/\V^{5/3}$. This is small enough not to destabilize
the mass of the volume modulus, $M_\ssV \simeq M_p/\V^{3/2}$.

\subsubsection*{Higher-dimensional contributions}

For scalar mass corrections, normally it is loops involving the
heaviest possible particles that are the most dangerous, so it
might be natural to expect the contributions from states having $M
\gg M_\KK$ to dominate the 4D estimates just obtained. In this
section we argue this not to be the case, with the dominant
contribution to modulus masses arising from loops involving states
at the KK scale.

The argument proceeds as in section \ref{sec:GenLoopEst}, with the
recognition that the effects of particles above the Kaluza-Klein
scale require the use of the higher-dimensional theory, where new
symmetries like extra-dimensional general covariance come to our
aid. In particular, since the largest mass scale in the higher
dimensional theory is $M_g$, or for Type IIB models the string
scale, $M_s \simeq g_s M_g$, the contributions of states this
massive are summarized by local string-loop and $\alpha'$
corrections to the 10D action.

However, any such contributions are strongly constrained by 10D
supersymmetry. The leading $\alpha'$ corrections arising at string
tree level are known, and first arise at $\O(\alpha'^3)$, with
four powers of the curvature. The volume dependence of this, and
of the other $\alpha'$ corrections that arise at next-to-leading
order were studied by refs.~\cite{LV}, with the conclusion that
these first contribute to the 4D scalar potential at order
\be
 \delta_{\alpha'} U \simeq \frac{W_0^2}{\V^3} \;
 \U_{\alpha'}(\varphi) \,,
\ee
where $\U_{\alpha'}(\varphi)$ denotes some function of the
dimensionless moduli (whose detailed form is not important for the
present purposes). Given their Planck-scale kinetic terms, this
contributes of order $\delta m \simeq \O(M_p/\V^{3/2})$ to the
low-energy moduli masses. Indeed, it is these $\alpha'$
corrections that lift the mass of the volume modulus in the first
place, showing why it is generically of order $M_p/\V^{3/2}$.

The $\V$-dependence of the leading string-loop effects has also
been analyzed \cite{BMP,CCQ,CCQTWO}, with the result that it
contributes at order
\be
 \delta_{\rm loop} U \simeq \frac{W_0^2}{\V^{10/3}} \;
 \U_{\rm loop} (\varphi) \,,
\ee
with $\U_{\rm loop}$ another function of the dimensionless moduli.
The resulting contribution is $\delta m \simeq M_p/\V^{5/3}$, in
agreement with the above estimates for the size of supersymmetric
cancellations in one-loop effects. This agreement indicates that
it is states having masses of order $M_\KK$ that provide the
dominant contribution to these string loops.

\subsection{Checks}

Large-volume string models have the advantage of having been
studied well enough that there are several nontrivial checks on
the above estimates.

\subsubsection*{Comparison with the 4D supergravity}

One check comes from comparing the above estimates of the size of
cancellations in one-loop supergravity amplitudes with the kinds
of corrections that are allowed to arise within the effective 4D
supergravity describing the moduli. In this supergravity the above
loop effects must appear as volume-dependent corrections to the
K\"ahler function, $K$, since this is the quantity that encodes
the effects of high-energy loops. The leading contributions to
$\delta K$ arising up to one loop has been estimated {\cite{BHK}}  , and has the
form
\be
 K \simeq - 2 \ln \V + \frac{k_1}{\V^{2/3}} + \frac{k_2}{\V} +
 \frac{k_3}{\V^{4/3}}\cdots \,,
\ee
where the $k_1$ term first arises at the level of one string loop
while $k_2$ contains the $\alpha'$ contributions at string tree
level, and the $k_3$ term gives the next corrections at one string
loop.

Recall that the leading contribution to the 4D potential varies as
$U \simeq W_0^2/\V^2$. It happens that because the $k_1$ term is
proportional to $\V^{-2/3}$, it completely drops out of the scalar
potential \cite{BMP,CCQ}. The $k_2$ term then gives the dominant
correction to $U$, contributing at order $1/\V^3$. Finally, the
$k_3$ term contributes a correction to $U$ that is of order
$1/\V^{10/3}$, all in agreement with the above estimates.

\subsubsection*{Comparison to explicit string calculations}

The explicit form of one-loop effects associated with Kaluza-Klein
(and string) modes are available for orbifolds and orientifolds of
toroidal compactifications. We here briefly summarize the strength
of these corrections and their interpretation in the low energy
effective field theory. This illustrates that the truncation to
the four dimensional effective field theory to be consistent and
the quantum corrections to the mass of the volume modulus
associated with Kaluza-Klein modes to be subleading in an
expansion in the inverse volume.

We focus on $N = 1$ orientifold ${{\mathbb{T}}}^{6} /
{\mathbb{Z}}_{2} \times {\mathbb{Z}}_{2}$ analyzed by Berg, Haack
and Kors in ref.~{\cite{BHK}}. The untwisted moduli in this model
are the three K\"ahler and complex structure moduli
$\{\rho^{\ssI}, U^{\ssI}\}$ axio-dilaton S, open string scalars
$A^{\ssI}$ associated with position of the $D3$ branes, the model
also has moduli associated with $D7$ brane positions which was set
to zero in {\cite{BHK}}.

The tree-level K\"ahler potential, found using the disc and sphere
string-worldsheet graphs, is given by
\begin{equation}
   K = - \ln ( S - \bar{S} )
 - \sum_{I=3}^{3} \ln \bigg[ (\rho^{\ssI}
 - \bar{\rho}^{\ssI})(U^{\ssI} - \bar{U}^{\ssI})
 - {1 \over 8 \pi} (A^{\ssI} - \bar{A}^{\ssI})^2 \bigg] \,,
\end{equation}
while the one-loop correction was computed in {\cite{BHK}}, and
found to be
\begin{equation}
  K^{(1)} = { 1 \over 256 \pi^6 } \sum_{I=1}^{3} \bigg[{ {
  {\mathcal{E}}_{2}^{D3} (A^{\ssI}, U^{\ssI}) } \over
 {(\rho^{\ssI} - \bar{\rho}^{\ssI})(S - \bar{S})}}  +
 \left. { {
 {\mathcal{E}}_{2}^{D7} (0, U^{\ssI}) } \over
 {(\rho^{\ssJ} - \bar{\rho}^{\ssJ})
 (\rho^{\ssK} - \bar{\rho}^{\ssK})}}
 \right|_{\ssK \neq \ssI \neq \ssJ} \bigg]
\end{equation}
where the superscripts $D3$ and $D7$ indicate contributions which
originate from open string diagrams with boundaries on these
branes. While the general expression for $\mathcal{E}_{2}^{D3}$
and $\mathcal{E}_{2}^{D7}$ are complicated, a symmetric choice of
the $D3$ brane positions (we refer the reader to {\cite{BHK}} for
details) the contribution from $D3$ branes vanishes and the
contribution from $D7$ branes is an Eisenstein series
\begin{equation}
 {\mathcal{E}}_{2}^{D7}(0,U)
 = \, 1920 \sum_{(n,m) \neq (0,0)} { { {\rm{Im}}(U)^{2}}
 \over {|n + U m|^{4} } }
\end{equation}

The indices $(m,n)$ can be interpreted as labelling the
Kaluza-Klien momenta of the exchanged particles in the open-string
loop diagrams. Note that the contribution from the higher KK modes
is suppressed by inverse powers of the KK momentum, as a result
the higher KK modes make only a small contribution. We note that
while the contribution of a single mode running in loop correction
to the mass would scale as the square of the mass ($n^2$),
cancelations due to supersymmetry lead to an effective scaling of
$n^{-4}$. We would like to emphasize that the resulting sum is
ultraviolet finite. One can interpret this as the restoration of
supersymmetry at the high scale.

The contribution of the one-loop K\"ahler potential to the scalar
potential scales as ${\cal{V}}^{-10/3}$ which indeed is subleading
in the large volume expansion. As mentioned earlier, this scaling
of the volume precisely matches the estimations of the size of
loop effects from KK modes in the low energy effective field
theory {\cite{CCQ}}.

The bottom line is that the combination of supersymmetry with the
$\V$-suppressed couplings coming from extra dimensions ensures the
stability of the masses of the moduli against radiative
corrections in large-volume string models.

\section{Scenarios}

Given the robustness of the light scalar masses, for
phenomenological purposes it is useful to see how large the above
masses are for various choices for the string scale, in order to
see precisely how light the relevant scalars can be. Since $\V$
varies exponentially sensitively with the parameters of the
modulus-stabilizing flux potential, it varies over many orders of
magnitude as potential parameters are changed only through factors
of order 10. It can therefore be regarded as a dial that can be
adjusted freely when exploring the model's implications.

\subsubsection*{Weak-scale gravitino mass}

One attractive choice is to take $\V \simeq 10^{15}$, in which
case $M_{3/2} \simeq M_p/\V \simeq 10^3$ GeV is of order the TeV
scale and $M_s \simeq M_g \simeq M_p/\V^{1/2} \simeq 10^{11}$ GeV,
corresponding to the intermediate-scale string \cite{ISS}.

In this case the generic moduli also have masses at the TeV scale,
while the volume modulus is interestingly light, being of order
$M_\ssV \simeq M_p/\V^{3/2} \simeq 10^{-3}$ GeV $\simeq 1$ MeV.
Even though very light, such a scalar would be very difficult to
detect, given its gravitational couplings to matter. It is not
light enough to mediate a measurable force competing with gravity,
and so is not constrained by the observational tests of gravity in
the lab or in astrophysics. Even though this contradicts the
standard lore that moduli masses are of the same order as the
gravitino mass after supersymmetry breaking, it actually makes the
cosmological moduli problem \cite{cmp}\  more severe since a
gravitationally coupled scalar field with mass $\sim 1$ MeV tends
to dominate the energy density of the universe through its
coherent oscillations after inflation. A low-energy mechanism to
dilute this field, such as a second period of inflation, may be
needed to avoid this problem. The physical implications of this
scenario are explored in more detail in \cite{astro}.

\subsubsection*{Intermediate scale gravitino mass and soft terms}

In \cite{ralph} a novel framework for supersymmetry breaking is
put forward in the context of the large volume scenario. Since the
main source of supersymmetry breaking is the $F$-term of the
volume modulus and, since the $4D$ supergravity is approximately
of the no-scale type, its contributions to soft terms can be
highly suppressed in powers of the volume, the $F$-terms of the
other K\"ahler moduli dominate the structure of the soft terms. If
the brane that hosts the standard model does not wrap the dominant
cycle for supersymmetry breaking, then the soft terms are
hierarchically suppressed with respect to the gravitino mass
$\Delta m\sim M_p/\V^q$ with $q={ 3\over 2},2$ depending on
potential cancellations. For $\Delta m\sim 1$ TeV, the gravitino
mass can be as high as $10^{10}$ GeV. Again we  have a situation
in which the soft terms, in particular the masses of the scalar
partners of the standard model fields, are much smaller than the
gravitino mass and loop corrections could in principle destabilize
these masses. However, the same arguments as before imply that
these loop corrections are at most proportional to
$M_{KK}M_{3/2}/M_p\simeq M_p/\V^{5/3}$ (see also
\cite{ralph,shanta}).

As a result squarks and sleptons much lighter than the gravitino
mass are naturally stable. A similar estimate can be made for the
other soft terms, such as $A$-terms and gaugino masses. This
implies that as long as the contributions to soft terms from the
volume modulus are suppressed, including anomaly mediated
contributions \cite{AM} (see however \cite{shanta0}), the gravitino mass can be hierarchically
heavier than the TeV scale. Notice that this ameliorates the
cosmological moduli problem mentioned above since the volume
modulus mass, which is of order $M_p/\V^{3/2}$, in this scenario
can be as heavy as $1$ TeV or heavier,  instead of being order MeV
as in the previous scenario.

\subsubsection*{Weak-scale strings and SUSY breaking on the brane}

The largest possible volume within this scenario that is
consistent with experience is $\V \simeq 10^{30}$, for which $M_s
\simeq M_p/\V^{1/2} \simeq 10^3$ GeV is at the weak scale, while
the moduli and gravitino are extremely light: $M_{\rm mod} \simeq
M_{3/2} \simeq M_p/\V \simeq 10^{-12}$ GeV $\simeq 10^{-3}$ eV. In
this case the volume modulus would appear to be astrophysically
relevant, since $M_\ssV \simeq M_p/\V^{3/2} \simeq 10^{18}$ eV.

More care is needed in this particular case, however, because the
supersymmetry breaking scale, $M_{3/2}$, is so very low. Because
it is so low, another source of breaking must be introduced for
the model to be viable, to accommodate the absence of evidence for
supersymmetry in accelerator experiments. The simplest such source
of supersymmetry breaking is hard breaking by anti-branes, with
all of the observed particles living on or near these branes so
that they are not approximately supersymmetric.

There are dangers to breaking supersymmetry in such a hard
fashion, however. In particular, for a viable scenario one must
check that the potential energy of the anti-brane --- which we
shall find is of order $M_s^4 \simeq M_p^4/\V^2$ --- does not
destabilize the LV vacuum.\footnote{We thank Joe Conlon for
emphasizing this point.} Notice that this is a more serious
problem than occurs if anti-branes are introduced to uplift the
previous cases so that their potentials are minimized to allow
flat spacetime on the branes. It is worse in this case because the
anti-brane must be the dominant source of SUSY breaking, and so
the value of its tension cannot be warped down to small values
without also making the mass splittings amongst supermultiplets
--- which are of order $M_s$ on the brane --- too small. In what
follows we assume that this issue has been dealt with, either
through inspired modelling or through fine-tuning.

Since the splitting between the masses of the bosons and fermions
localized on the brane is of the order $M_s$, it is loops
involving these brane states that are the most dangerous
corrections to $M_\ssV$ in this picture. We now argue that these
loops of brane states generate mass corrections of order $M_p/\V$,
and so in this particular case lift the volume modulus to be
comparable in mass to the masses of the other moduli.

The estimate begins with the Born-Infield action on the brane and
the ten dimensional Einstein action in the bulk
\begin{equation}
   S =  M^{8}_g\int \exd^{10} x \sqrt{g_{10}} {\cal R }_{10}
   - T_{3} \int \exd^{4} x \sqrt{ \det [G_{\alpha \beta}] }
\label{act}
\end{equation}
where $T_3 \simeq \O(M_s^4)$ is the brane tension and $G_{\alpha
\beta}$ is the pullback of the ten dimensional metric to the brane
world volume. Neglecting powers of the string coupling, we take
the 10D Planck scale of order the string scale: $M_g \simeq M_s$.

To carry out the dimensional reduction we take, as before, the ten
dimensional metric to be of the form
\begin{equation}
      \exd s_{10}^{2} = \omega e^{-6u(x)} g_{\mu \nu}
      \exd x^{\mu} \exd x^{\nu}
      + e^{2u(x)} g_{mn} \exd y^{m} \exd y^{n},
\end{equation}
where, as before, $\omega = (M_p/M_s)^2$, $u(x)$ is the volume
modulus and $g_{mn}$ is the metric of a Calabi-Yau of unit volume
in string units. Working with static embedding coordinates for the
brane the effective action for the volume modulus $u(x)$ and the
transverse scalars $y^{m}$ is
\begin{eqnarray}
 - 24 M_{p}^{2} \int \exd^{4} x \, \sqrt{-g_4}\;
 \partial_{\mu} u \partial^{\mu} u
 -   \hat T_{3} \int \exd^{4} x \sqrt{-g_{4}} e^{-12 u(x)}
      \sqrt{ \det[( \delta^{\alpha}_{\phantom{\alpha}\beta}
      + e^{8u(x)} \partial^{\alpha} y^{m}
      \partial_{\beta} y^{n} g_{mn})]} \nn\\
\label{effact}
\end{eqnarray}
where $M_{p}^{2} $ is the four dimensional Planck mass and $\hat
T_3 = \omega^2 T_3 \simeq \O(M_p^4)$.

Next we make field redefinitions which bring their kinetic terms
to the canonical form, $u(x) =  u_{0} + { \ell(x) / 4 \sqrt{3}
M_{p} } $ and $ y^{m} = { e^{2u_{0} }  \phi^{m} / \sqrt{\hat
T_{3}} }$, where $u_{0}$ is the v.e.v. of the field $u(x)$ and
related to the volume of the compactification in string units by $
e^{6u_{0}} = {\cal{V}}$. This brings (\ref{effact}) to the form
\begin{eqnarray}
 - { 1 \over 2} \int \exd^{4} x \sqrt{- g_{4}} \partial_{\mu} \ell
 \partial^{\mu} \ell - {\hat T_{3} \over {{\cal{V}}^{2}} }
 \int \exd^{4} x \sqrt{-g_{4}} e^{- \sqrt{3} \ell / M_{p} }
  \sqrt{ \det\left[ \left( \delta^{\alpha}_{\phantom{\alpha}     \beta}
  +  { {{\cal{V}}^{2}} \over \hat T_{3}}
  {{ e^{2\ell /  \sqrt{3} M_{p}} }  }
  \partial^{\alpha} \phi^{m} \partial_{\beta}
  \phi^{n} g_{mn}\right) \right]} \nn\\
\label{effactwo}
\end{eqnarray}
We note that the derivative expansion  for interactions involving
the fields $\phi^{m}$ is controlled by the scale $ \Lambda = ( {
{\hat T_{3} / {\cal{V}}^{2}}} )^{1/4}$. At the two-derivative
level interactions between $\ell$ and $\phi^{m}$ are Planck
suppressed, the leading order term being
\begin{equation}
  { 1 \over M_{p} } \int \exd^{4}x \sqrt{-g_{4}} \;  g_{mn}
  \partial^{\alpha} \phi^{m} \partial_{\alpha} \phi^{n}{ \ell(x) },
\end{equation}

The graph for the one loop correction to the mass of the volume
modulus involving this interaction has two inverse powers of
$M_{p}$ from the interaction vertices. The relevant integral over
the virtual momenta has four powers of momenta from the
integration measure, four from the two interaction vertices and
four inverse powers from the propagators. This leads to a loop
correction to the mass squared of the volume modulus of the order
of $\Lambda^4 / M^2_{p} $.  The above estimate in fact holds for
all contributions arising from virtual loops of the fields
$\phi^{m}$. This is most easily seen from dimensional analysis.
The associated graph involves two legs of the field $\ell(x)$, and
since the field always appears in the combination of ${\ell /M_{p}
}$ this leads to two factors of $1/M_p$. Next note that for
processes involving loops of $\phi^{m}$, all the vertices and loop
momenta are powers of ${\hat T_{3} / {\cal{V}}}$. This implies
that the mass correction is of the form $\delta m^{2} \propto (
{\hat T_{3} / {\cal{V}}^2} )^{p} { / M_{p}^{2} } $, where on
dimensional grounds the value of $p$ must be $p=1$. Since $\hat
T_{3} \sim \O(M_{p}^{4})$, the scale $\Lambda \sim { M_{p} /
{{\cal{V} }}^{1/2} } \sim M_{s} $ is of order the string scale, so
the size of the mass correction is
\begin{equation}
 \delta m  \propto  {\Lambda^2 \over M_p}
 \simeq {M^{2}_{s} \over M_{p}}
 \simeq { M_{p} \over {\cal{V} } } \,.
\end{equation}
This implies the mass of the volume modulus is naturally of order
$M_{\rm mod} \sim M_{3/2} \sim M_p/\V$ when supersymmetry is badly
broken on a brane, removing the hierarchy between the volume
modulus and other moduli.

If the string scale is $\sim 1$ TeV, then all of these light
scalar masses are of order $M_p/\V \simeq 10^{- 3}$ eV, making
them just on the edge of relevance to laboratory tests of the
gravitational force law. Because it is so close, corrections can
be important, and -- as shown in ref.~{\cite{astro}} -- the
couplings between brane matter and bulk fields give an additional
logarithmic suppression to the volume modulus mass, which then
scales as $M_p /(\V \ln \mathcal{V})$. If this correction pushes
the volume modulus into the range probed by terrestrial
fifth-force experiments, it could make the effects of this field
detectable. The existence of such a scalar as a robust consequence
of weak-scale string models within the large-volume picture
provides additional motivation for more detailed calculations of
its mass and properties in the presence of anti-branes.

TeV string scenarios could have spectacular experimental
implications at LHC (see for instance \cite{dieter}, and
references therein), so it is of great interest that they might be
viable within the large volume scenario for which control over
issues like modulus stabilization and supersymmetry breaking
allows a detailed prediction of the low-energy spectrum. The
existence within this spectrum of a very light volume modulus was
the main obstacle to serious model building, so the existence of
mass generation by explicit breaking of supersymmetry on the brane
is of particular interest.\footnote{We thank I. Antoniadis, G.
Dvali and D. L\"ust for useful conversations on this issue.}

\section{Conclusions}

In this paper we show that extra dimensions and supersymmetry can
combine to to protect scalar masses from quantum effects more
efficiently than either can do by itself. Supersymmetry by itself
would not necessarily protect masses lighter than the gravitino
mass and extra dimensions in principle need not protect scalar
masses lighter than the KK scale. But both together allow scalar
masses to be hierarchically smaller than both the KK and gravitino
masses. We call this kind of unusual scalar hierarchy {\it
\"uber-natural}.

New mechanisms for keeping scalar masses naturally light are
interesting because they are both rare and potentially very
useful, both for applications to particle physics and cosmology.
Because light scalars tend to have many observable consequences,
their existence can help identify those models to which the
ever-improving tests of general relativity on laboratory and
astrophysical scales are sensitive. From a model builder's
perspective, light scalars are also useful because they can cause
problems, such as the cosmological moduli problem, and thereby
focus attention on those models that can deal with these problems.

Our power-counting estimates show that \"uber-naturally light
scalars ultimately remain light because their masses and
interactions are systematically suppressed by powers of $1/\V =
M_g^2/M_p^2$, showing that they are light because the gravity
scale is lower than the 4D Planck scale. But because scalars are
potentially so UV sensitive, to properly establish that their
masses are naturally small requires knowing the UV completion of
gravity, within a framework that includes an understanding of
modulus stabilization and supersymmetry breaking. The possibility
of studying this quantitatively has only recently become possible,
within the large-volume vacua of Type IIB flux compactifications
in string theory.

The large volume scenario predicts clear hierarchies in the
low-energy spectrum of scalar states, within a framework of
calculational control that allows us to be explicit about the size
of quantum corrections. It also allows several sub-scenarios,
depending on the size of the volume and the location of the
standard model within the extra dimensions. We discuss a few of
the preliminary implications of our results for several of these.
\begin{itemize}
\item We find that soft supersymmetry breaking terms ($\Delta m
\propto 1/\V^{3/2}, 1/\V^{2}$) much smaller than the gravitino
mass $M_{3/2} \propto 1/\V$ can be stable against quantum
corrections, since these are smaller or equal to $1/\V^{5/3}$.
This is important for the stability of those recent models where
the soft supersymmetry breaking relevant to weak scale particle
physics is parametrically small compared with the gravitino mass,
such as happens when the main contribution to supersymmetry
breaking comes from cycles different from the cycle wrapped by the
standard model brane.
\item We find that TeV scale string models not only can be
obtained from the large volume scenario, but that their main
potential observational obstacle -- the existence of an extremely
light volume modulus -- might not be such a problem, making them
much more appealing. The volume modulus need not be a problem
because the relatively large mass corrections it receives from the
strong breaking of supersymmetry on the brane that is required in
such models.
\end{itemize}

More generally, {\em \"uber-naturalness} provides the mechanism
that underlies many of the attractive features of LV models that
have proven valuable for phenomenology. For instance, LV models
ultimately bring the news of supersymmetry to ordinary particles
through a form of gravity mediation, yet avoid the normal pitfalls
(such as with flavour-changing neutral currents) of gravity
mediation for low-energy phenomenology \cite{MirrorMed}. {\em
\"Uber-naturalness} provides the framework for the stability of
this process against loop corrections.

Similarly, inflationary models have been constructed within the LV
scenario with the inflaton being a volume modulus \cite{LVI}\ or
another K\"ahler modulus \cite{KMI}, with the latter achieving a
slow roll by virtue of the inflaton taking large field values,
rather than requiring a tuning of parameters in the potential.
These scenarios profit from the {\em \"uber-natural} protection of
the potential within the LV picture, indicating that
extra-dimensional symmetries like general covariance can provide
an alternative to global shift symmetries (see for instance
\cite{shiftsym}) for addressing the $\eta$-problem.

Further implications can well be envisaged. In particular, the
suppressed corrections to the masses of light moduli may be useful
for cosmology by providing new and better models of
inflation. Perhaps new models of dark energy could become
possible with this new naturalness concept in mind. We hope that
our results will at least serve to stimulate the search, in as
explicit a manner as possible, for further suppression mechanisms
for scalar fields in theories with supersymmetric extra
dimensions.

\section*{Acknowledgements}

We wish to thank Ignatios Antoniadis, Joe Conlon, Matt Dolan, Gia
Dvali, Sven Krippendorf, Louis Leblond, Dieter L\"ust and Sarah
Shandera for useful discussions. Various combinations of us are
grateful to the Center for Theoretical Cosmology in Cambridge,
Perimeter Institute, McMaster University, the Cook's Branch
Workshop in Houston and the Abdus Salam International Center for
Theoretical Physics, for their support and provision of pleasant
environs where some of this work was done. We also thank
Eyjafjallajokull for helping to provide us with undivided time to
complete this paper. AM was supported by the EU through the Seventh
Framework Programme and the STFC. CB's research was supported in
part by funds from the Natural Sciences and Engineering Research
Council (NSERC) of Canada. Research at the Perimeter Institute is
supported in part by the Government of Canada through Industry
Canada, and by the Province of Ontario through the Ministry of
Research and Information (MRI).

\appendix

\section{Extra-dimensional kinematics vs KK sums}

In this appendix we discuss the  loops correction in the higher
dimensional theory from the perspective of the lower dimensional
theory.  Consider a massless scalar field in D dimensions with a
quartic coupling of strength of the order of the higher
dimensional cutoff
\ba
   -{ \cal{L}_{D} } = - {1 \over 2} (\partial \Phi)^2
   + {g_4 \over 24} \Phi^4 \,,
\ea
with
\ba
  g_{4} = \bigg( {1 \over \Lambda} \bigg)^{(D-4)} \,.
\ea
Upon dimensional reduction on a $d$-torus
\ba
  \exd s^{2} = \eta_{\mu \nu} \exd x^{\mu} \exd x^{\nu}
  + L^{2} \delta_{mn} \exd y^{m} \exd y^{n}
\ea
and canonical normalization this gives the lower dimensional
action for the KK modes of the form
\ba
  -{ \cal{L} }_{4}   =  - {1 \over 2} \sum_{n_{i}}
  \bigg[(\partial \phi_{n_{i}})^2 + {1 \over 2}
  m^2_{n_{i}}  \phi^{2}_{n_{i}}\bigg]
  + {g_4 \over  L^{d}} \sum_{n_{i},n_{j},n_{k},n_{l}}
  c_{n_{i} n_{j} n_{k} n_{l}} \phi_{n_{i}}
  \phi_{n_{j}} \phi_{n_{k}} \phi_{n_{l}} \,,
\ea
where $i,j,k,l = 1 .. d$, $ m^2_{n_{i}} = { 1 \over L^{2} }
\sum_{i} n_{i}^2 $ and the interaction coefficients $c_{n_{i}
n_{j} n_{k} n_{l}}$ are of order one if the KK charge associated
with the vertex is vanishing.

Now let us consider the loops in the above theory, the  loop
contribution due to a KK of mass $m_{n_{i}}$ is
\ba
 \delta m_{n_{i}}^{2}  \approx { g_{4} \over L^{d} }
 m_{n_{i}}^{2} \,.
\ea
In order to estimate the effect of the entire KK tower we sum the
loop contributions of all KK modes up to the scale $\Lambda$ i.e
we restrict the KK momenta to $|n_{i}| < N_{\rm max} = \Lambda L$.
This gives
\ba
 \sum_{n_{i}} \delta m_{n_{i}}^{2} \approx
 { g_{4} \over L^{2+d} } (\Lambda L)^{2+d} = \Lambda^{2} \,,
\label{lsum}
\ea
in agreement with the discussion in section \ref{sec:GenLoopEst}.

We note that the estimate assumed no cancellations in the sum over
the loop contributions  in (\ref{lsum}), as is appropiate for a
scalar. But, as emphasized in section \ref{sec:GenLoopEst} if one
considers gravity in higher dimensions, the restrictions on the
form of higher dimensional action imposed by general covariance
necessarily implies  cancellations between the loop contributions
of various particles in the lower dimensional theory. Such
cancellations can lower the size of loop corrections to scales
parametrically below the cut off scale of the higher dimensional
theory.

\end{document}